\documentclass[iop, apj, twocolumn, tighten, twocolappendix, numberedappendix]{emulateapj}

\usepackage[dvipsnames]{xcolor}
\usepackage{amsmath}
\usepackage{amssymb}
\usepackage{graphicx}
\usepackage{natbib}
\usepackage{xspace}
\usepackage{nicefrac}
\newcommand\myshade{85}
\usepackage[T1]{fontenc}
\usepackage{lmodern}

\usepackage[backref,breaklinks,colorlinks,citecolor=blue]{hyperref}        
\usepackage[all]{hypcap}

\colorlet{mycitecolor}{Turquoise}
\colorlet{mylinkcolor}{Turquoise}

\hypersetup{
 citecolor = mycitecolor!\myshade!black,
 linkcolor = mylinkcolor!\myshade!black,
 colorlinks = true,
}

\def\apjl{Astrophysical Journal}             

\def\apjs{ApJS}              
\def\aap{A\&A}                

\newcommand{\bh}[1]{$\mathrm{BH}_{#1}$}

\def\be{\begin{equation}}
\def\ee{\end{equation}}
\def\ba{\begin{eqnarray}}
\def\ea{\end{eqnarray}}

\begin{document}

\title{$N-$body dynamics of Intermediate mass-ratio inspirals in globular clusters}

\def\addBham{School of Physics and Astronomy, University of Birmingham, Edgbaston, Birmingham B15 2TT, United Kingdom}

\def\addNU{Center for Interdisciplinary Exploration and Research in Astrophysics (CIERA)\\ \& Dept. of Physics and Astronomy, 2145 Sheridan Rd, Evanston, IL 60208, USA}

\author{Carl-Johan Haster}
\email{cjhaster@star.sr.bham.ac.uk}
\affiliation{\addBham}
\affiliation{\addNU}

\author{Fabio Antonini}
\affiliation{\addNU}

\author{Vicky Kalogera}
\affiliation{\addNU}

\author{Ilya Mandel}
\affiliation{\addBham}

\begin{abstract}
The intermediate mass-ratio inspiral of a stellar compact remnant into an intermediate mass black hole (IMBH) can produce a gravitational wave (GW) signal that is potentially detectable by current ground-based GW detectors (e.g., Advanced LIGO) as well as by planned space-based interferometers (e.g., eLISA). 
Here, we present results from a direct integration of the post-Newtonian $N$-body equations of motion describing stellar clusters containing an IMBH and a population of stellar-mass black holes (BHs) and solar mass stars. 
We take particular care to simulate the dynamics closest to the IMBH, including post-Newtonian effects up to order $2.5$. 
Our simulations show that the IMBH readily forms a binary with a BH companion. 
This binary is gradually hardened by transient 3-body or 4-body encounters, leading to frequent substitutions of the BH companion, while the binary's eccentricity experiences large amplitude oscillations due to the Lidov-Kozai resonance. 
We also demonstrate suppression of these resonances by the relativistic precession of the binary orbit.
We find an intermediate mass-ratio inspiral in one of the 12 cluster models we evolved for $\sim 100$ Myr.
This cluster hosts a $100 M_\odot$ IMBH embedded in a population of 32 $10M_\odot$ BH and 32,000 $1M_\odot$ stars. 
At the end of the simulation, after $\sim 100$ Myr of evolution, the IMBH merges with a BH companion. 
The IMBH--BH binary inspiral starts in the eLISA frequency window ($\gtrsim 1\rm mHz$) when the binary reaches an eccentricity $1-e\simeq 10^{-3}$. After $\simeq 10^5$ years the binary moves into the LIGO frequency band with a negligible eccentricity.
We comment on the implications for GW searches, with a possible detection within the next decade.
\end{abstract}

\section{Introduction}
\label{sec:intro}

Intermediate mass black holes (IMBHs) are conjectured to occupy the mass range between stellar-mass black holes (BHs), with masses $\lesssim 100 M_\odot$, and supermassive black holes with masses $\gtrsim 10^6 M_\odot$ \citep[see][for a review]{MillerColbert:2004}. 
While the existence of some IMBH candidates in dwarf spheroidal galaxies has been conjectured by extending the $M$--$\sigma$ relation \citep{GrahamScott:2013} \citep[but see][]{MaccaroneServillat:2008}, dynamical measurements of IMBHs in the few-hundred solar-mass range are extremely challenging \citep[e.g.,][]{Pasquato:2016}. 
The best evidence for such lower mass IMBHs (with mass $\sim 100 M_\odot$) could come from ultraluminous X-ray sources \citep[but see][]{Colbert:2008}; for example, \citep{Pasham:2014} have claimed a mass of $\sim 400$ $M_\odot$ for M82 X-1 from quasi-periodic oscillations,
while a mass around $10^4 M_\odot$ has been suggested for the brightest ultraluminous X-ray source HLX-1 \citep[e.g.,][]{Farrell:2009,Davis:2011,Godet:2014}, but these dynamical measurements alone can not provide conclusive proof for the existence of IMBHs.

If these lower-mass IMBHs reside in globular clusters, they will play an important role in cluster dynamics \citep[e.g.][]{Trenti:2006,Umbreit:2013, Konstantinidis:2013,Leigh:2014,Macleod:2016}. 
Of particular interest to our study is the likely tendency of IMBHs to dynamically form compact binaries with other compact remnants \citep[e.g.][]{Taniguchi:2000,Miller:2002,MillerHamilton:2002a,AmaroSeoane:2006,Brown:2007,Mandel:2008,AmaroSeoaneSantamaria:2009,Mapelli:2010,Mapelli:2016}. 
Generally, these analyses find that the IMBH readily captures a binary companion. 
The binary is subsequently hardened through a sequence of 3-body and 4-body interactions, occasionally with substitutions which make a black hole (BH) of a few tens of solar masses the most likely IMBH companion, and possible Lidov-Kozai (LK) resonances \citep{Lidov:1962,Kozai:1962} if hierarchical triples are formed. 
Eventually, the IMBH--BH binary merges through the radiation of gravitational waves, emitting a signal that is potentially detectable by the Advanced LIGO ground-based GW detectors \citep{AdvLIGO,ratesdoc,Smith:2013,Haster:2015}. 

Previous simulations of globular clusters with IMBH coalescences have generally simplified the interactions in order to avoid excessive computational cost. 
For example, \citet{Gultekin:2004} considered a series of individual Newtonian interactions interspersed with orbital evolution through GW emission. 
\citet{Mandel:2008} carried out analytical estimates of the hardening sequence to obtain the intermediate mass-ratio merger timescale.
\citet{Leigh:2014} simulated the entire cluster with a mixture of analytical and numerical $N$-body analytical calculations, while \citet{Macleod:2016} focused their $N$-body investigation on tidal disruptions of stars by the IMBH as well as merger events. 
We note that in the previous literature effects of pN terms are either not accounted for \citep{Leigh:2014}, or included only at the 2.5pN level \citep{Samsing:2014,Macleod:2016}. 
In this paper we show a clear example in which lower order pN terms play a fundamental role in the dynamics.
More specifically, an essential element that differs between the relativistic and non-relativistic dynamics turns out to be the 1pN precession of the periapsis.

We introduce our numerical method and the simulation setup in \autoref{sec:code}. 
We describe our simulation results in \autoref{sec:results}. 
We discuss the results, including the detectability of GWs from intermediate mass-ratio coalescences, in \autoref{sec:discuss}.

\section{Simulations}
\label{sec:code}

The $N$-body systems considered here consist of a massive particle, representing an IMBH, and two additional lower-mass species representing $10M_\odot$ compact remnants and $1M_\odot$ stars.
Integrations of the $N$-body equations of motion were carried out using the direct summation $N$-body code phiGRAPEch \citep{Harfst:2008}.
This code incorporates Mikkola's algorithmic chain regularization scheme including post-Newtonian terms of order 1pN, 2pN and 2.5pN \citep[AR-CHAIN,][]{Mikkola:2008}.
Velocity dependent forces were included using the generalized midpoint method described by \citet{Mikkola:2006}.
The algorithm produces exact trajectories for Newtonian two-body motion and regular results for strong encounters involving arbitrary numbers of bodies. 
Particles moving beyond the ``chain radius'' ($r_{chain}$) were advanced using a fourth-order integrator with forces computed on GPUs using the Sapporo library \citep{Gaburov:2009}. 
The chain particles were influenced by the global cluster dynamics through the particles in a perturber region, within a radius $r_{perturb}$ from the IMBH.
phiGRAPEch is an ideal tool for the study of the dynamics of IMBHs in star clusters because it allows to study with extremely high precision the joint effect of 1pN, 2pN and 2.5pN terms and their interplay with Newtonian perturbations to the motion.

We performed 12 simulations all initialized as a King model with $W_0=7$, no primordial binaries, containing two mass species (BHs and stars) with a relative mass ratio of $10:1$, and assuming that the total mass in BHs is $1\%$ of the total cluster mass. 
Finally, an initially stationary IMBH was placed at the center of the cluster.
The simulations were performed with the number of particles $N \subset \{32768, 65536\}$ and the mass of the IMBH, $M \subset \{50, 100, 200\}\, M_\odot$ and a cluster virial radius $r_v$ of $3.5$ pc. 
For $N = 32768$ and all three IMBH masses, simulations with $r_v \subset \{0.35, 1.0\}$ pc were also performed.
The inclusion of high-order pN terms fixes the physical scale of the cluster, thus removing the conventional freedom for rescaling simulations in cluster size and density. 

We observe the IMBH forming a binary with a BH within $\lesssim 20$Myr in every simulated cluster. 
Only in one cluster $(N=32768, M=100 M_\odot, r_v=3.5$pc) we observe a merger within the simulated time ($\simeq 100\rm Myr$). 
While the result of the entire set of simulations will be presented in a future paper, in what follows we will focus on describing the detailed dynamics of the one cluster producing the merger event. 
We note in passing that as the main focus of this study is the dynamical formation and evolution of binaries, and higher order $N$-tuples, with the IMBH as the primary companion, all cluster particles are solely characterized by their mass and no stellar evolution is included in these simulations.

\section{Results}\label{sec:results}

\begin{figure*}[t]\center
    \includegraphics[width=0.75\textwidth,keepaspectratio]{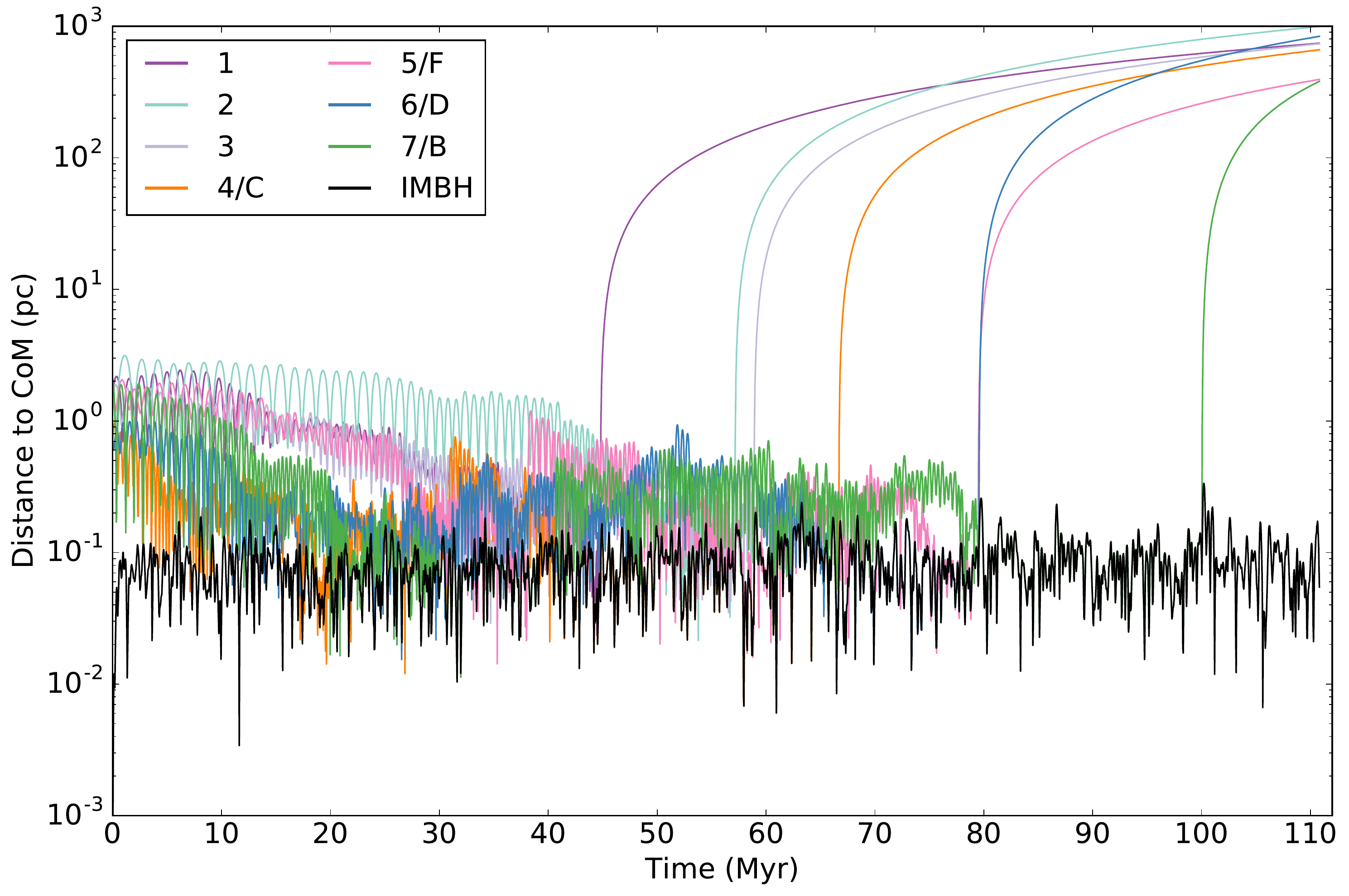}%
    \caption{Time evolution of the distance of the IMBH (in black) to the center of mass (CoM) of the entire cluster. 
		 The IMBH wanders throughout the simulation within a central region extending to $\lesssim 0.1\rm pc$ around the cluster CoM. 
		Also shown (in color) are the BHs that were ejected from the cluster and the corresponding time of ejection. 
		 BHs for which we have assigned both a numerical and alphabetical index were bound to the IMBH before being ejected from the cluster. 
		The evolution of the orbits of these BHs is also shown in \autoref{fig:sub_trip}.} 
\label{fig:wander_eject}
\end{figure*}

The simulated globular cluster was initialised with the IMBH at rest at the center while the remaining stars and BHs follows a King model.
\autoref{fig:wander_eject} shows the position of the IMBH and a subset of BH particles, and their subsequent movement within the cluster, relative to the center of mass of the entire cluster.
This subset of the BH population were those that were ejected from the cluster during the simulation.
Although the IMBH is initially at rest at the cluster center of mass, it quickly experiences significant Brownian motion within a sphere of radius $\sim 0.1$pc around the center of mass. 
The typical distance wandered by the IMBH in the core is larger than the radius of influence of the IMBH.

\begin{figure*}[t]\center
      \includegraphics[width=0.75\textwidth,keepaspectratio]{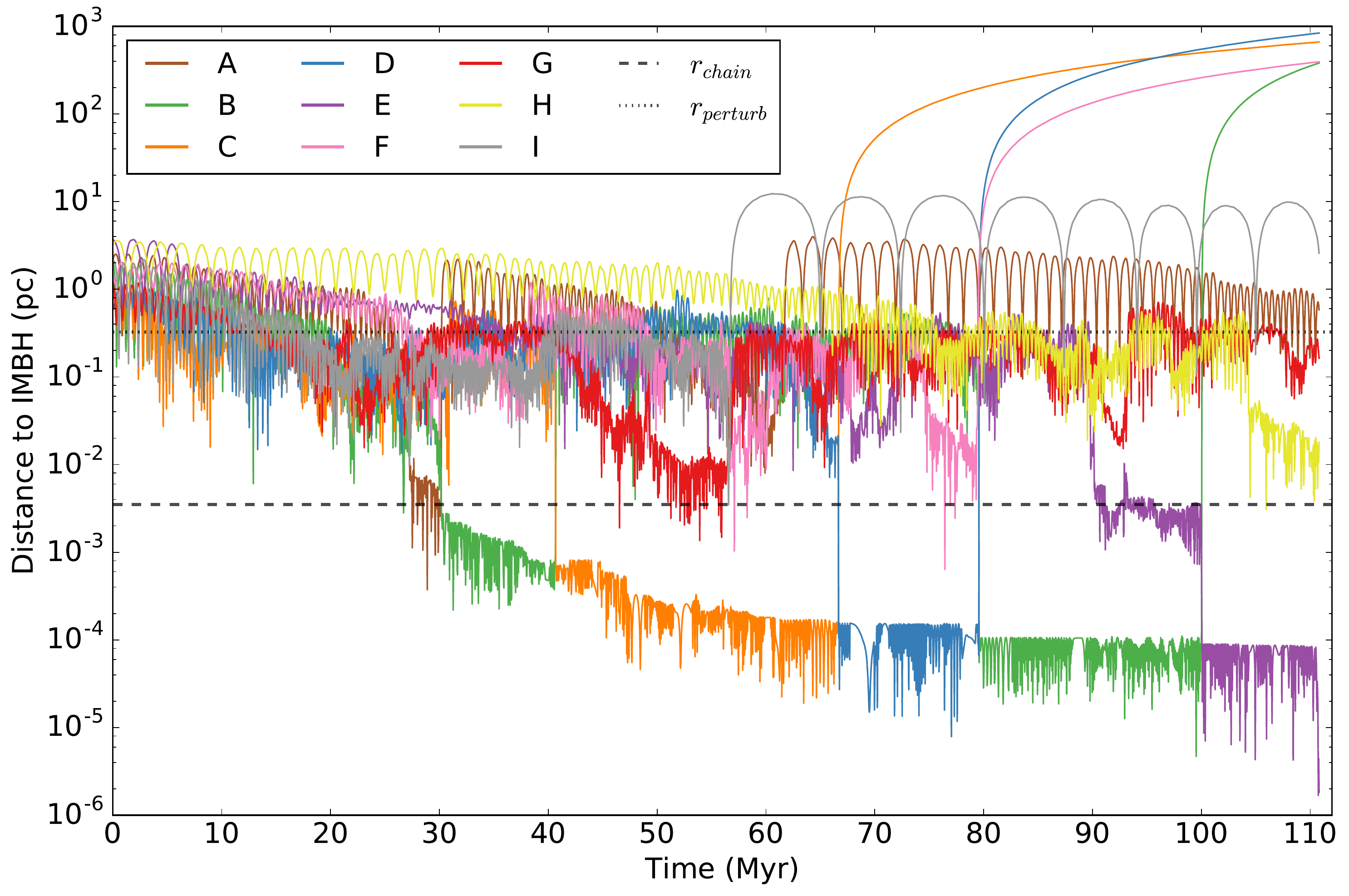}%
    \caption{Distance to the IMBH versus time for BHs which at any point in the simulation came within 4 mpc of the IMBH.
		Within the first 3 Myrs the IMBH acquires a binary companion, at first a stellar particle which is substituted for a BH companion at $\sim 25$ Myrs, in this case \bh{A}, forming a wide binary. 
		Through interactions with other objects embedded in the cluster potential this binary is hardened. 
		The IMBH--BH binary undergoes many companion substitutions, often while in hierarchical $N$-tuples, resulting in both later recaptures (\bh{B}) and ejections from the cluster (\bh{C,D,B}). 
		The dashed black line marks the transition distance $r_{chain}$ below which the dynamics are treated by AR-CHAIN under the gravitational influence of all perturbing particles within the region represented by the dotted black line $(r_{perturb})$. 
		} 
\label{fig:sub_trip}
\end{figure*}

Shifting the focus from the global dynamical behaviour within the cluster, \autoref{fig:sub_trip} displays the time evolution of the relative distance to the IMBH of those BHs which experienced close encounters with the IMBH at some point of the simulation. 
In this figure we see that while the IMBH is interacting only weakly with its surroundings at the start of the simulation, after $\sim 3$ Myrs it forms a wide binary with a stellar particle, and after $\sim 25$ Myrs the binary companions are BHs, consistent with the expected mass segregation in this cluster.
By comparing the ejected BHs between \autoref{fig:wander_eject} and \autoref{fig:sub_trip} it is clear that after the first few ejected BHs (which were driven by their initially relatively high kinetic energy and interactions with other cluster members) and following the formation of the IMBH--BH binary, all subsequent ejections are driven by interactions with the IMBH--BH binary.
These interactions lead to the frequent substitution of the IMBH binary companion, with three out of the five observed substitution events leading to the former companion being ejected from the cluster. 
The remaining two were returned to the cluster BH population, where one BH (\bh{B} in \autoref{fig:wander_eject} and \autoref{fig:sub_trip}) was later recaptured by the IMBH. 
\autoref{fig:sub_trip} also shows the transient 3-body interactions, such as the ejection of \bh{F}, and their effectiveness in the continued hardening of the IMBH--BH binary. 

\begin{figure*}[t]\center
     \includegraphics[width=0.75\textwidth,keepaspectratio]{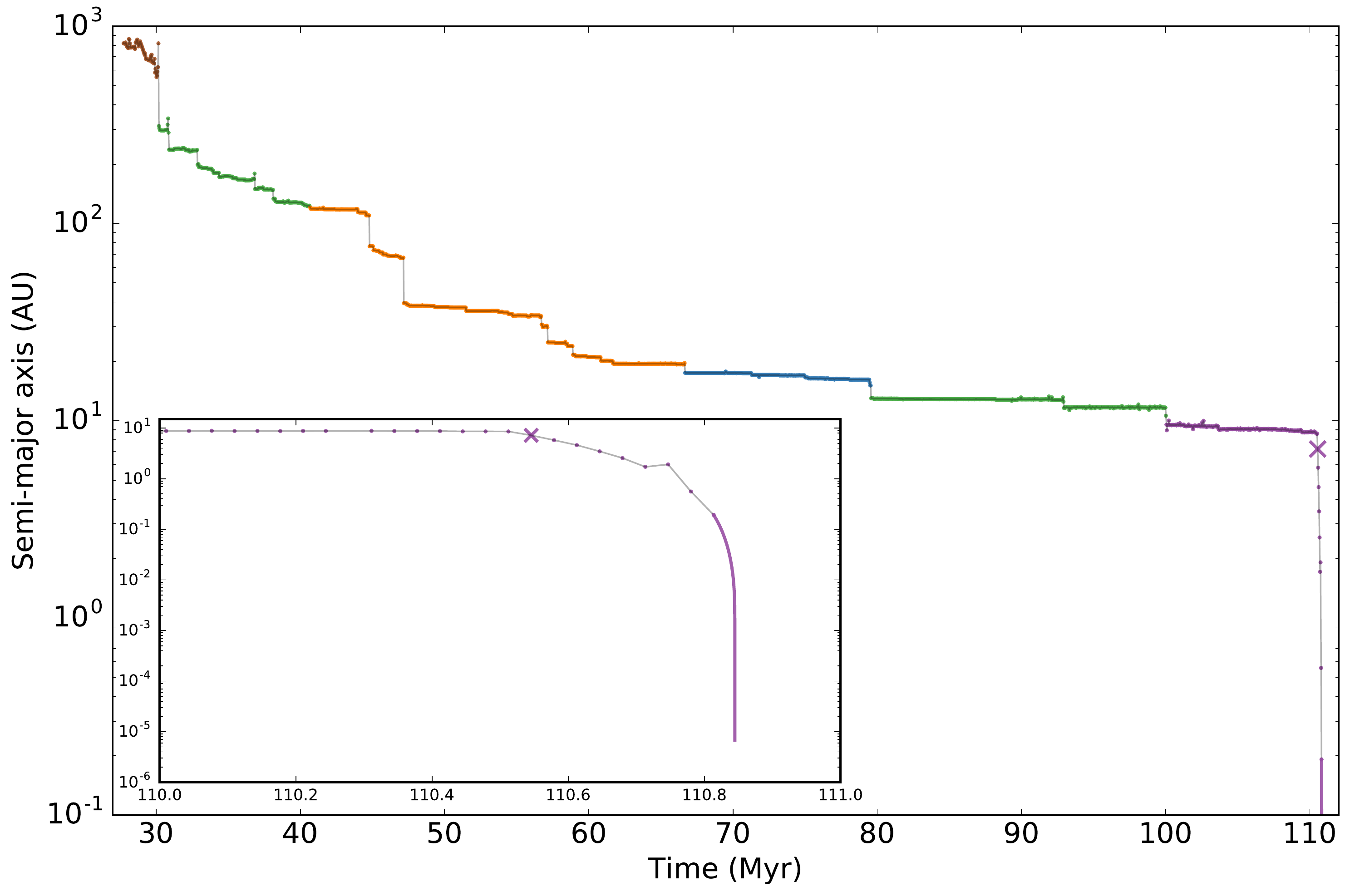}%
    \caption{A stable binary consisting of an IMBH and another BH will harden over time and lose energy to its surroundings, manifested by the shrinking of the binary's semi-major axis. 
		The colors match \autoref{fig:sub_trip} to highlight the substitution of the binary companions. 
		At the end of the simulation the IMBH, through external perturbation by \bh{H}, is set on a trajectory towards a merger with its binary companion \bh{E} while still inside the cluster. 
		The point where GW emission becomes dominant in the orbital evolution (see \autoref{eq:GW_dominating}) is marked by a purple $\times$. 
		At the end of the simulation the binary orbit is evolved until merger according to \citet{Peters:1964} marked by the solid purple line, this is further highlighted in the inset figure showing the last Myr before merger.} 
\label{fig:sma}
\end{figure*}

The time evolution of this binary is most clearly visualized in terms of its orbital parameters where \autoref{fig:sma} shows the IMBH--BH binary semi-major axis.\footnote{The semi-major axes and eccentricities were computed using the post-Newtonian formalism given in Equation (3.6) of \citet{Damour:1985}.} 
Once the IMBH captures a stellar-mass BH compation, the IMBH--BH binary is hardened by 3-body interactions.
The hardening of the binary is clearly visible in \autoref{fig:sma} as the semi-major axis of the IMBH--BH binary decreases monotonically, with the jumps in semi-major axis being signs of energetic 3-body interactions. 
While \autoref{fig:sub_trip} only shows the BH interactions, there are also a multitude of stellar transient passes carrying energy away from the IMBH--BH binary.

Gravitationally focused interactions with which approach the hard IMBH binary within its semi-major axis $a_i$ happen on a typical timescale \citep[e.g.,][]{Mandel:2008}
\begin{eqnarray}
\tau_\textrm{3-body} \simeq 5 \times 10^7 \left(\frac{100 M_\odot}{M_b}\right) \left(\frac{v}{10\textrm{km s}^{-1}}\right) \nonumber \\
\left(\frac{10^{5.5}\textrm{pc}^{-3}}{n}\right) \left(\frac{0.05\rm AU}{a_i}\right) \textrm{yr}\ ,
\label{eq:tau_3body}
\end{eqnarray} 
where $M_b$ is the binary's mass (dominated by the IMBH), $v$ is the velocity dispersion in the cluster and $n$ is the number density of stars and BHs in the cluster center.
The binary hardens through 3-body interactions on the typical timescale \citep[e.g.,][]{Quinlan:1996,Mandel:2008}
\begin{equation}
\tau_\textrm{harden} \simeq \frac{22}{\pi} \frac{M_b}{m_*} \tau_\textrm{3-body}\ ,
\label{eq:tau_harden}
\end{equation}
where $m_*$ is the interloper mass. 
Following \citet{Antonini:2016c} we define $a_{ej}$ as the binary semi-major axis below which an three-body interaction will cause the binary to be ejected from the cluster:
\begin{equation}
a_{ej} \simeq 0.2G\mu \frac{m_*}{M_b+m_*}\frac{m_*}{M_b}\frac{1}{v_{esc}^2}
\label{eq:a_ej}
\end{equation}
where $\mu = Mm/(M+m)$ for $M$ is the IMBH mass, $m$ is the mass of its BH companion and $v_{esc}$ is the escape velocity from the core of the cluster.
As the binary hardens, after $\simeq 110\rm Myr$ the time to the next interaction drops below the GW driven merger timescale which, in the limit of large binary eccentricities, is approximated by \citep{Peters:1964} 
\begin{equation}
\tau_\textrm{merge} \simeq 3\times10^{7}
\left(10^5M_\odot^3 \over M^2m \right)  \left(a_i\over 0.05\rm AU \right)^4(1-e_i^2)^{7/2}\rm{yr} \ ,
\label{eq:tau_merge}
\end{equation}
where $e_i$ is the IMBH--BH binary eccentricity. 
The semi-major axis $a_{\rm GW}$ at which the evolution of the binary starts to be dominated by GW radiation, and no further significant 3-body interactions are expected, can be found by setting 
\begin{equation}
\tau_\textrm{merge}=\tau_\textrm{3-body}
\end{equation}
which gives
\begin{equation}
\label{eq:GW_dominating}
a_{\rm GW}\simeq \frac{0.12}{\left(1-e_i^2\right)^{7/10}} \rm AU 
\end{equation}
with $M=100 M_\odot$, $m = 10 M_\odot$, $n = 5 \times 10^3 \textrm{pc}^{-3}$ and $v = 10\textrm{km s}^{-1}$ reflecting the cluster center at the time of the merger onset seen in \autoref{fig:sma}.
At separations below $a_{\rm GW}$ the evolution of the binary is dominated by energy loss due to GW emission and for $a_{\rm GW} > a_{ej}$ the merger will occur before the binary will be ejected from a three-body interaction. 

\citet{Mandel:2008} computed the total time to IMBH--BH coalescence by summing the hardening time to the last interaction with the subsequent merger timescale under the assumption that the last interaction is likely to leave the binary with an eccentricity of $\simeq 0.98$ \citep{Gultekin:2006}. 
However, as shown in what follows, even higher eccentricities can be reached during the complex 3-body interactions, possibly reducing the merger timescale \citep{Antonini:2014,Samsing:2014}.

\begin{figure*}\center
      \includegraphics[width=0.75\textwidth,keepaspectratio]{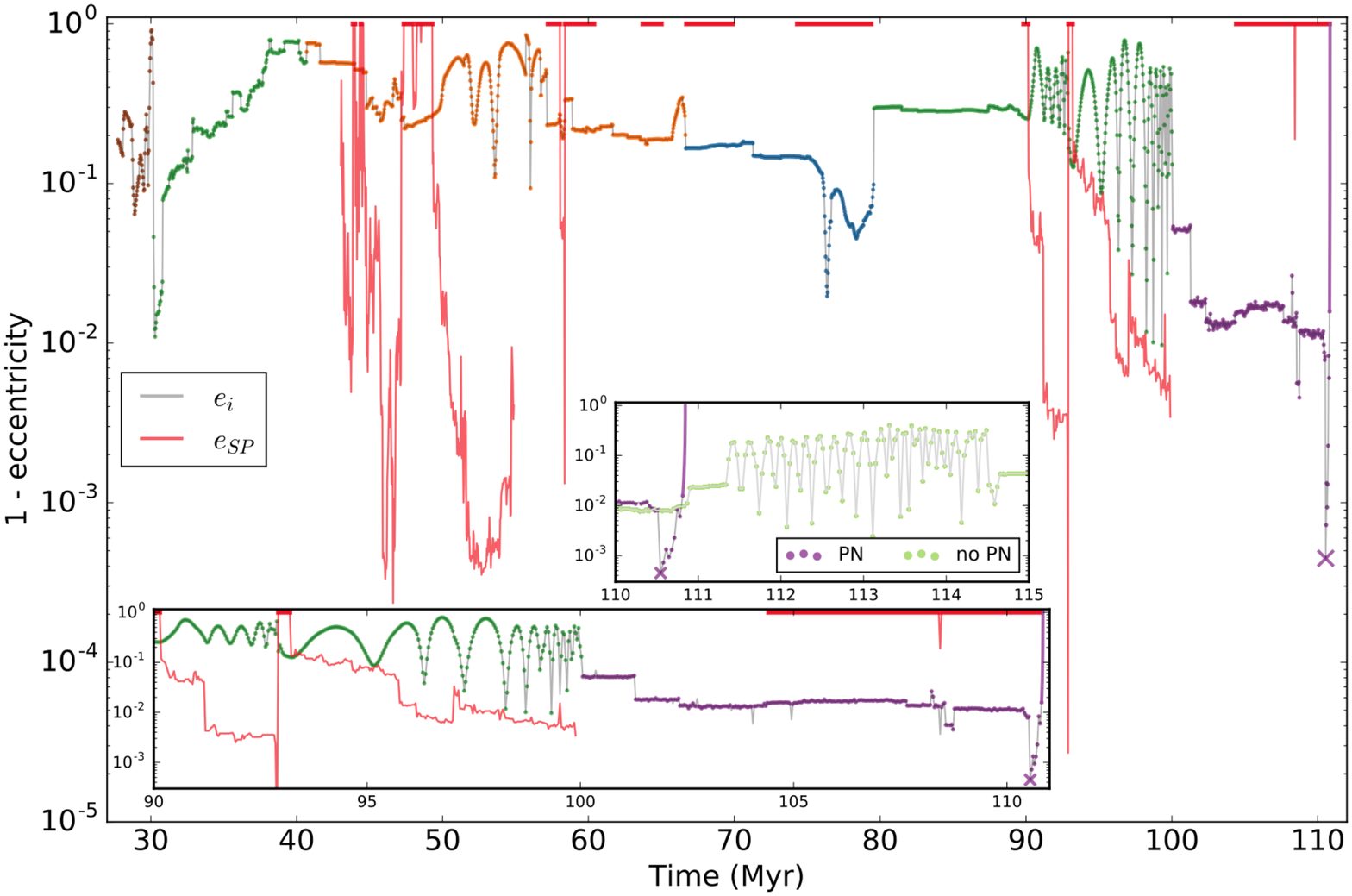}
      \includegraphics[width=0.75\textwidth,keepaspectratio]{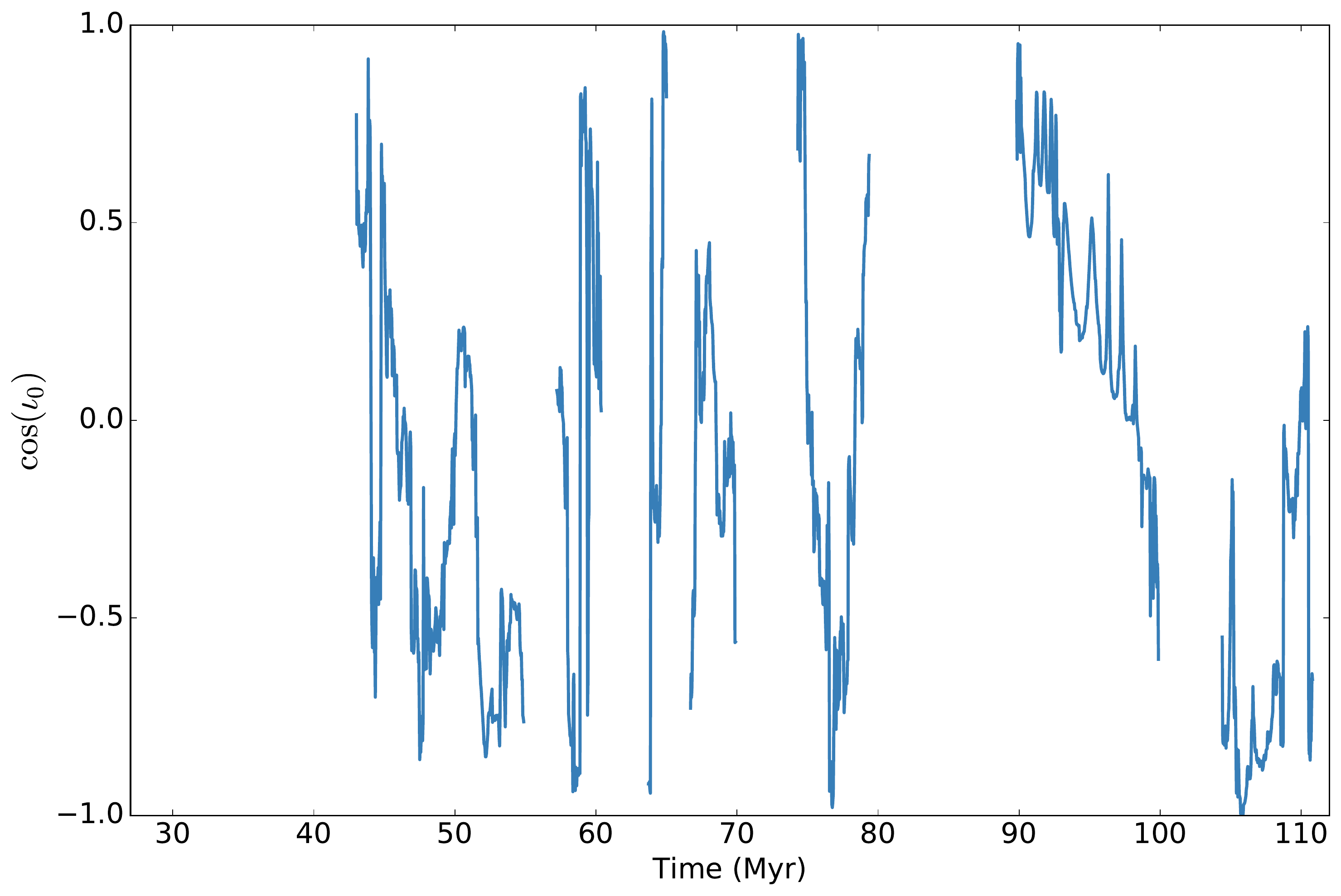}
    \caption{Evolution of the IMBH--BH binary eccentricity (top).
		The colors of $e_i$ match those of \autoref{fig:sub_trip} to highlight the many substitutions of the binary companions. 
		The figure shows clear evidence for both transient 3-body interactions as well as longer timescale LK oscillations. 
		For the majority of the binary's presence in a hierarchical $N$-tuple LK effects are suppressed by the SP of the inner binary orbit. 
		The precession is represented here as an effective eccentricity boundary $e_{SP}$ above which LK oscillations are suppressed (corresponding to \textit{below} the $e_{SP}$ line in this figure). 
		The bottom inset shows the last $\sim 20$ Myr exhibiting LK oscillations, bounded by $e_{SP}$, in a quadruple and later a triple BH system ended by the substitution of the IMBH binary companion. 
		This last binary configuration is frozen at high $e_i$, suppressed by SP and later merged.
		The point where the binary evolution is dominated by GW emission is marked by a purple $\times$. 
		At the end of the simulation the binary orbit is evolved using the formula of \citet{Peters:1964} marked by the solid purple line. 
		To further highlight the importance of the post-Newtonian dynamics the upper inset also includes a simulation (presented in light green) started at $\sim 1$ Myr before the observed merger, but using only Newtonian dynamics. 
		The clear eccentricity oscillations in a triple system where SP would have completely suppressed LK provide evidence for the importance of pN dynamics.
		Also shown in the bottom panel is the inclination $\iota_0$ between the inner and outer orbits for the times when the IMBH exists in a bound triple system.
		}
\label{fig:ecc}
\end{figure*}

\autoref{fig:ecc} shows the evolution of the binary eccentricity (top panel) and inclination of outer to inner binary when the binary is part of a triple system ($\iota_0$; bottom panel).
The complex dynamical structure of the surroundings of the binary, including both stellar and BH interactions, is evident in \autoref{fig:ecc} where large amplitude variations of the IMBH--BH binary eccentricity are observed.
The evolution of $e_i$ and $\iota_0$ is driven both by transient passes and by longer duration LK and post-Newtonian effects. 

The eccentricity oscillations in hierarchical triple systems can potentially drive up the eccentricity of the inner binary to very high values, possibly leading to faster GW driven mergers \citep{MillerHamilton:2002a,Aarseth:2012,Antonini:2014}. 
The timescale for a full oscillation in eccentricity is given as
\be
\label{eq:LK_period}
\mathrm{T}_{LK} \simeq \frac{P_i}{2\pi} \frac{M+m}{m_o} \left(\frac{a_o}{a_i}\right)^3 (1 -e_o^2)^{3/2}
\ee
where $a_i$, $a_o$ are the semi-major axes of the inner and outer binary respectively (within the hierarchical triple system), $e_o$ is the eccentricity of the outer orbit, $m_0$ is the mass of the tertiary BH, and $P_i = 2\pi \sqrt{a_i^3/G(M+m)}$ is the orbital period of the inner binary \citep{Holman:1997}. 
We also define here a dimensionless angular momentum, as the angular momentum of the binary divided by the angular momentum of a circular orbit with the same semi-major axis: $\ell_i = \sqrt{1- e_i^2}$; this is a useful quantity when discussing LK oscillations, as they do not affect the orbital energy.
The timescale over which the inner binary changes the value of its angular momentum by order of itself is then \citep{Bode:2014,Antonini:2016a}
\be
\tau_{LK} \equiv \left|\frac{1}{\ell_i}\frac{d\ell_i}{dt}\right|_{LK}^{-1} \simeq \mathrm{T}_{LK} \sqrt{1 - e_i^2} \ ,
\label{eq:tau_LK}
\ee
directly related to the period of the LK oscillation.

At the quadrupole level of approximation and in the test particle limit, for an orbit librating around the argument of periapsis $\omega_i =\pi/2$ the maximum ($\ell_+$) and minimum ($\ell_-$) angular momenta during a LK cycle are related through the equation \citep[e.g.,][]{Merritt:2013}
\be 
\ell_+\ell_- = \sqrt{\frac{5}{3}}\ell_z .
\label{eq:emax1}
\ee
In the previous expression $\ell_z=\ell_i\cos \iota_0$ is a conserved quantity for an initial orbital inclination $\iota_0$ in the quadrupolar limit . 
From \autoref{eq:emax1}, and from the conservation of $\ell_z$ one finds that the maximum eccentricity that can be attained during a LK cycle is simply $e_{max} = \sqrt{1 - (5/3)\cos^2\iota_0}$ \citep{Innanen:1997}.

Post-Newtonian corrections to the orbital dynamics can affect the binary on similar timescales as $\tau_{LK}$, where the most prominent effect would be the 1pN Schwarzschild precession (SP) of the argument of periapsis $\omega_i$. 
To lowest order, the timescale associated with SP is 
\be
\tau_{SP} \equiv \left|\frac{1}{\pi}\frac{d\omega_i}{dt}\right|_{SP}^{-1} \simeq \frac{P_i}{6}\frac{a_i}{r_g}(1 - e_i^2)
\label{eq:tau_sp}
\ee
with $r_g = G(M+m)/c^2$. 
When SP is considered, \autoref{eq:emax1} becomes \citep{Antonini:2016b}:
\be
\label{eq:lib}
\ell_+^2\ell_-^2={\frac{5}{3}}\ell_z^2 +\frac{k}{3}\left(\frac{\ell_+-\ell_-}{\ell_+^2-\ell_-^2}\right)\ell_+\ell_-,
\ee
with 
\be
k=8\frac{M}{m}\frac{r_ga_o^3}{a_i^4}\left(1-e_o^2\right)^{3/2}\ .
\ee
This shows that SP effects can suppress the phase space available for libration for systems with $\tau_{SP} < \tau_{LK}$, reducing the maximum eccentricity attained during a LK cycle. 
In fact, from \autoref{eq:lib}, given that the second term on the right hand side of the equation is always positive, we see that for a given $\ell_+$ SP will lead to an increase of $\ell_-$.

By setting $\tau_{LK} = \tau_{SP}$ we find the critical angular momentum
\be
\ell_{SP} \sim \frac{r_g}{a_i}\frac{M+m}{m_o}\left(\frac{a_o}{a_i}\right)^3
\label{eq:l_SP}
\ee
which in turn can be represented as an eccentricity boundary $e_{SP} = \sqrt{1- \ell_{SP}^2}$ to the eccentricities within reach of LK oscillations.
SP will dominate the orbital evolution of the inner binary at eccentricities larger than $e_{SP}$, thus quenching the possibility of eccentricity oscillations caused by LK resonance.
When $\ell_{SP} \geq 1$, then SP will dominate over the torque from the outer tertiary BH for any value of $e_i$ and LK oscillations are expected to be fully suppressed. 
In \autoref{fig:ecc} we show when this happens using a solid red line at $\ell_{SP}=1$. 
As expected, no LK oscillations occur when $\ell_{SP} \gtrsim 1$.
In \autoref{fig:ecc} the $e_{SP}$ boundary is shown only when there is a hierarchical triple system present, with the IMBH--BH binary at its center. 
During the periods of active LK oscillations, for example between $\sim 90-100$ Myr, it is clear that the eccentricity of the IMBH--BH binary never exceeds the $e_{SP}$ boundary. 
This is further evidence that SP plays a fundamental role in the dynamical evolution of the IMBH--BH binary in our simulations. 
The detailed interaction between LK and SP dynamical effects is also discussed by \citet{Naoz:2013} who find that, assuming Newtonian dynamics to octupolar order with an added 1pN (only) correction term, SP in hierarchical triples can excite eccentricity rather than suppress it for $\tau_{SP} \sim \tau_{LK}$.
While qualitatively similar behaviour can be observed in our simulation, it is difficult to distinguish effects like this from other mechanisms subdominant to the LK oscillations (e.g. the hierarchical mass ratio configuration, stellar interlopers and the perturbing cluster potential) without further investigation.

In addition to LK suppression from relativistic precession, the presence of strong Newtonian precession, induced by the IMBH--BH binary existing within a dynamical cluster, would have similar effects on the binary orbital evolution. We find the classical precession to be negligible compared to SP for the periods when the IMBH--BH binary is in a hierarchical triple, and thus have no effect on the LK suppression caused by precession of the IMBH--BH orbit.

Between $\sim 90-93$ Myr we find that the IMBH--BH binary is part of a hierarchical quadruple BH (IMBH, \bh{B}, \bh{E}, \bh{F}), with a resolved two-level LK oscillation. 
As discussed by \citet{Hamers:2015}, as the individual $\tau_{LK}$ for the two LK systems are comparable, this induces complex LK oscillations in the IMBH--BH binary, further enhancing the transfer of angular momentum away from it. 
This is most clearly exemplified by the eccentricity: the expected maximum eccentricity $e_{max} \simeq 0.3$ from $\iota_0 = 43.1^{\circ}$ at $90$ Myr is substantially smaller than the eccentricities achieved during the existence of the quadruple BH. 
It is also interesting to note that the two LK timescales associated with the quadruple are both below the corresponding $\tau_{SP}$ for the inner binary as well as the empirical timescale for the precession of $\omega_i$ induced by the presence of the quadruple within the stellar cluster.
Eventually at $\simeq 93$ Myr the quadruple system is disrupted by the removal of the outermost BH. 
At $93-100\rm Myr$ of evolution the eccentricity of the IMBH--BH binary clearly undergoes large-amplitude LK oscillations as expected given the high mutual inclination of the outer to inner orbit $(\iota_0 = 78.5^{\circ})$ at this time.

While the observed oscillations in eccentricity and inclination of the IMBH triple system show all signs of being caused by the LK mechanism, it is important to keep in mind that ``pure'' LK oscillations assume an isolated 3-body system. 
Here we are able to observe (for the first time) this mechanism acting on a triple system embedded in a dynamically evolving stellar cluster while also accounting for relativistic corrections to the motion.

Also interesting are the regions where $e_{SP} = 0$, or equivalently $\ell_{SP} \geq 1$. 
Here the SP is dominating the IMBH--BH binary to such a degree that no eccentricity oscillations driven by LK are possible. 
We find that the inclination of the IMBH--BH binary relative to the outer BH orbit evolves stochastically, attaining at times values near $40^{\circ}$ which based on \autoref{eq:emax1} should lead to LK eccentricity oscillations with a period $\simeq 5\times 10^6$ years following \autoref{eq:LK_period}.

\begin{figure*}[t]\center
   \includegraphics[width=0.75\textwidth,keepaspectratio]{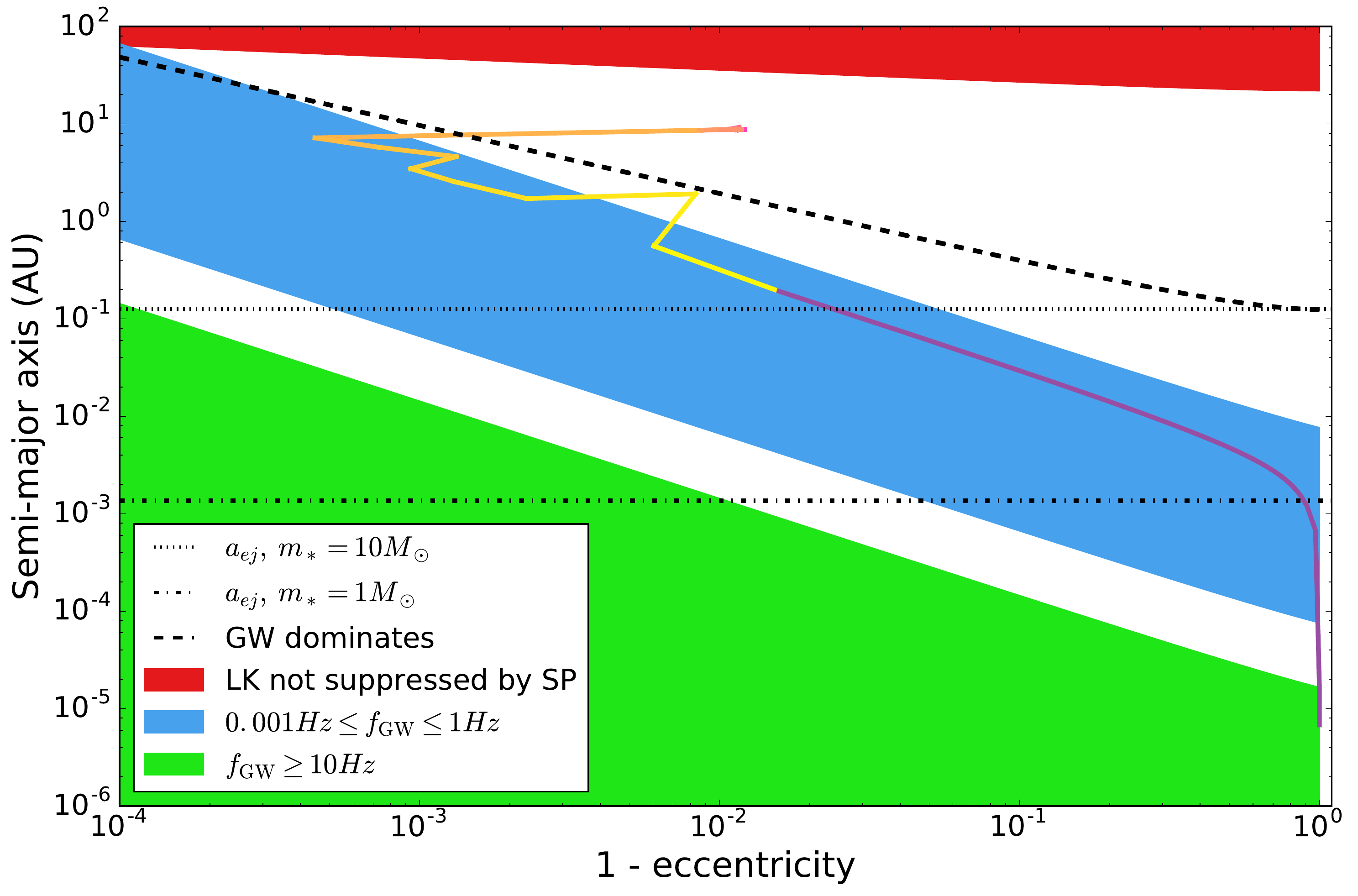}%
  \caption{As the IMBH--BH binary evolves during its final Myr (time increases from red to yellow) it is frozen at very high eccentricities due to the suppression of LK oscillations by the SP of the IMBH--BH binary.
		The red region indicates where this suppression would be absent. 
		The presence of a stable triple system, as indicated in \autoref{fig:sub_trip}, causes perturbations of the IMBH orbit from both the tertiary \bh{H} and additional stellar interlopers. 
		One of the stellar perturbations brings the three objects so close together that the IMBH--BH binary orbital evolution becomes dominated by emission of GWs. 
		Additionally, this ejects the stellar interloper at a velocity $\sim 120 \textrm{km s}^{-1}$.
		The limiting semi-major axes below which a three-body interaction with an interloper of mass $m_*$ will eject the binary are shown as dotted and dash-dotted lines following \autoref{eq:a_ej}.
		GW emission dominates below the dashed black line, given by \autoref{eq:GW_dominating}; in this regime merger through the emission of GWs will occur before the next 3-body interaction can significantly alter the IMBH--BH binary eccentricity, and thus its evolutionary timescale (c.f. the interloper which initiated the merger trajectory).
		Much of the GW dominated evolution occurs at GW frequencies observable by eLISA, as indicated by the blue region. 
		As the IMBH--BH binary evolves along its merger trajectory there are still a small number of minor three body encounters with stellar interlopers passing within a few semi-major axes of the binary CoM; these interactions are the cause of the ``spikes'' visible in the merger trajectory.
		These interactions are consistent with the timescales given in \autoref{eq:tau_harden} and \autoref{eq:tau_merge}, which predict that the last interaction before merger should occur when this system has a semi-major axis of $\sim 1$ AU. 
		At the end of the simulation the binary's orbit is evolved to merger, within the Advanced LIGO sensitive band (marked by the green region), according to \citet{Peters:1964}, as shown by the purple line.} 
\label{fig:merger}
\end{figure*}

Contrary to this, during the last $\sim 10$ Myr the IMBH--BH binary appears to be ``frozen'' at high $e_i$ with SP suppressing any eccentricity reduction apart from the higher pN-order emission of GWs. 
After a strong interaction with a stellar interloper, which in turn is ejected from the cluster at $\sim 120 \textrm{km s}^{-1}$, the loss of energy and angular momentum through GWs determines the ultimate fate of the binary, leading to its merger $\sim 300,000$ years later (see \autoref{fig:merger}).

During the initial GW-dominated phase the binary experiences a small number of 3-body interactions with stellar interlopers, which are the cause of the ``spikes'' visible in \autoref{fig:merger}.
However, these interactions do not disrupt the binary inspiral and since they occur at $a_i > a_{ej}$, shown in \autoref{fig:merger} for both possible interloper masses and $v{esc} = 9.9 \textrm{km s}^{-1}$, the interaction does not eject the binary from the cluster.

A highly eccentric binary emits a broad spectrum of gravitational radiation during each periapsis passage. 
We identify the frequency of the harmonic containing the maximal gravitational radiation as \citep{Wen:2003}:
\begin{equation}
f_{\rm GW}=\frac{\sqrt{G(M+m)}}{\pi}
{{\left(1+e_i\right)^{1.1954}}\over \left[a_i\left(1-e_i^2\right)\right]^{1.5}}\ ;
\end{equation}
this is the GW frequency plotted in \autoref{fig:merger}.

The binary spends $\simeq 300,000$ years in the eLISA sensitive frequency window, which spans $(0.001 \rm{Hz} \lesssim f_{\rm GW} \lesssim 1 \rm{Hz})$.
Meanwhile, the last $6$ seconds of the inspiral, followed by the merger and subsequent ringdown of the resulting IMBH, occur in theGW spectrum observable by Advanced LIGO ($f_{\rm GW}\gtrsim 10\rm{Hz}$). 
As suggested by \citet{AmaroSeoaneSantamaria:2009} for IMBH--IMBH binaries and \citet{GW150914astro:2016, Sesana:2016} for binary BH systems similar to the detected GW150914 \citep{GW150914:2016}, this type of IMBH--BH coalescences represents a class of GW sources potentially observable in both space- and ground-based detectors, providing an opportunity for long-term detailed studies of both their formation environments and probes of general relativity itself. 
We discuss the near-term prospects of detecting such binary mergers with Advanced LIGO in the following section.

To further verify the importance of the inclusion of the pN effects in our models, the simulation was restarted $\sim 1$ Myr before the merger, removing all pN terms from the equations of motion. 
As shown by the light green samples in the upper inset of \autoref{fig:ecc}, removing pN terms results in eccentricity oscillations without, of course, a GW induced merger.
During these oscillations the IMBH remains in a bound triple system (IMBH, \bh{E}, \bh{H}) in which SP would have been the dominant dynamical factor, completely removing the possibility for LK oscillations.

In addition, we performed one simulation, also started $\sim 1$ Myr before the merger, where the chain regularization was disabled. 
In this simulation the surrounding $N$-body integrator could not accurately follow the very hard IMBH--BH binary; this had the effect of significantly slowing down the simulation while no longer adhering to the conservation of total energy within the cluster. 
This loss of $\frac{\Delta E_{}}{E_{}} \sim 0.01$ per time-step $\Delta t=67 \rm kyr$ accumulates as the simulation progresses, to be compared with $\frac{\Delta E_{}}{E_{}} \sim 10^{-4}$ per time-step when including the chain regularization. 
These results further demonstrate the importance of using a high accuracy integrator like AR-CHAIN in order to study the evolution of IMBHs in cluster simulations.

\section{Discussion}
\label{sec:discuss}

Gravitational waves from intermediate mass-ratio coalescences are observable with both the advanced network of ground-based detectors \citep{AdvLIGO, AdvVirgo} and a future space-based GW detector \citep{eLISA:2013}.   
The observed binary inspiral is in the eLISA sensitivity band throughout the circularizing phase. 
However, detection and parameter estimation at very high eccentricities could prove problematic without high-accuracy eccentric templates for matched filtering the bursts of radiation expected during the few periapsis passages over the lifetime of a space-borne mission \citep{AmaroSeoane:2007,Porter:2010,Key:2011,Berry:2013}. 
Both detection and parameter estimation would be more amenable to existing techniques later in the orbital evolution \citep{Huerta:2014,Tiwari:2016,Tanay:2016,Moore:2016,Nishizawa:2016}, for $e_i \lesssim 0.1$ in \autoref{fig:merger}, and with only ten years from $e_i = 0.1$ until merger for this system a co-observing campaign together with ground-based detectors, where the system is effectively fully circularized, would be possible.

As discussed by \citet{Sesana:2016,Vitale:2016} in relation to binary BHs similar to GW150914, the extended observation in eLISA would provide excellent constraints on the binary masses, sky position and coalescence time with the observational gap of $1 Hz \le f_{\rm GW} \le 10 Hz$ between eLISA and Advanced LIGO only spanning $\sim 1$ hour. 
This advance information would allow for optimization of the ground-based detector network, both in terms of active tuning of the detector sensitivity, operational scheduling and the analysis pipelines, as well as pre-pointing of electromagnetic follow-up telescopes\footnote{No electromagnetic counterpart is expected from the merger of an IMBH--BH binary in the standard scenario \citep[e.g.,][]{Lyutikov:2016}, but see \citet{Fermi:2016}}.
For the remainder of this section we will focus on detectability and rates for ground-based detectors alone, primarily motivated by the lack of a space-based detector for at least the next decade.

\begin{figure*}\center
 \hspace{0.33in} \includegraphics[width=0.8\textwidth,keepaspectratio]{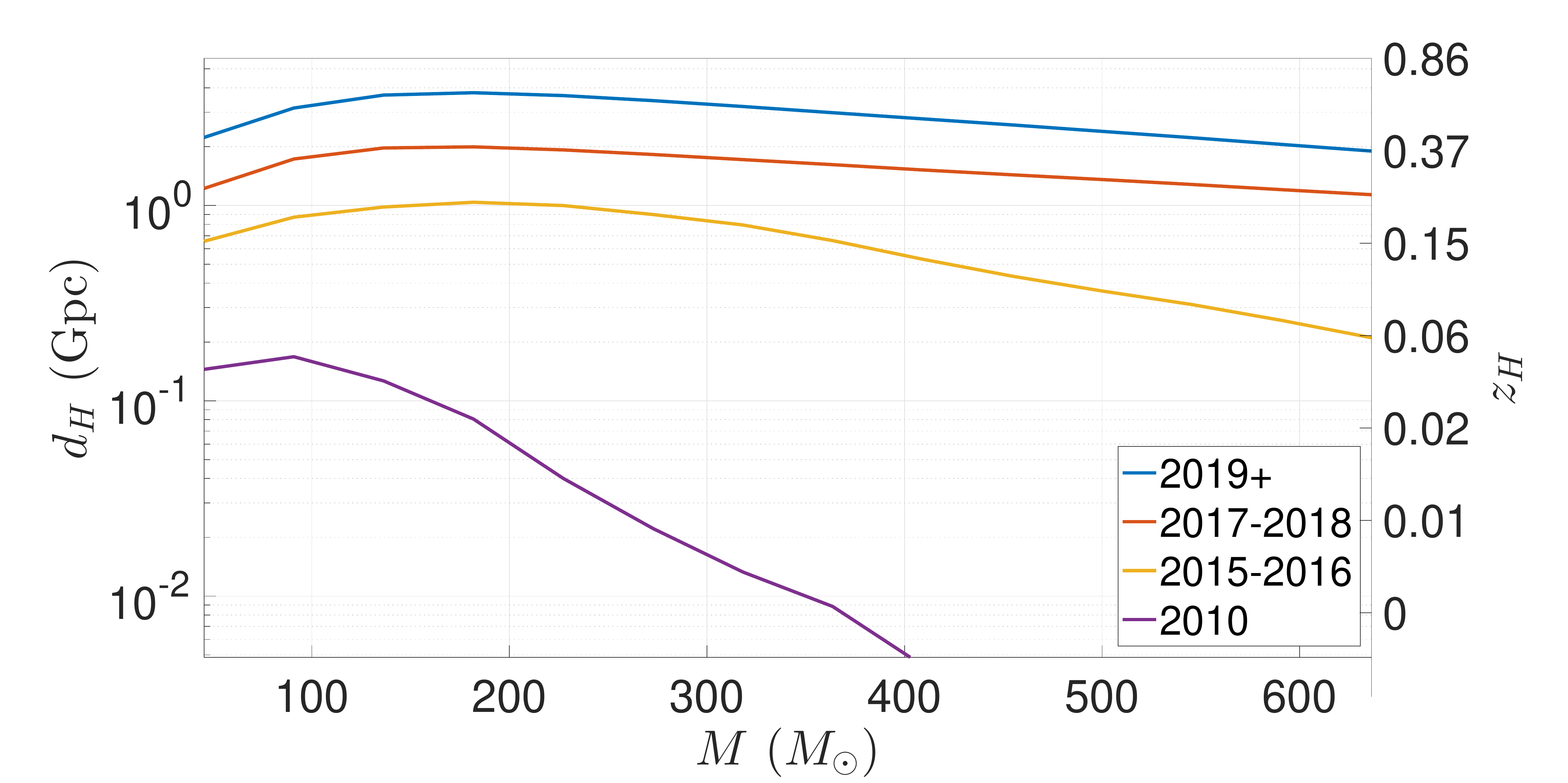}
 \includegraphics[width=0.68\textwidth,keepaspectratio]{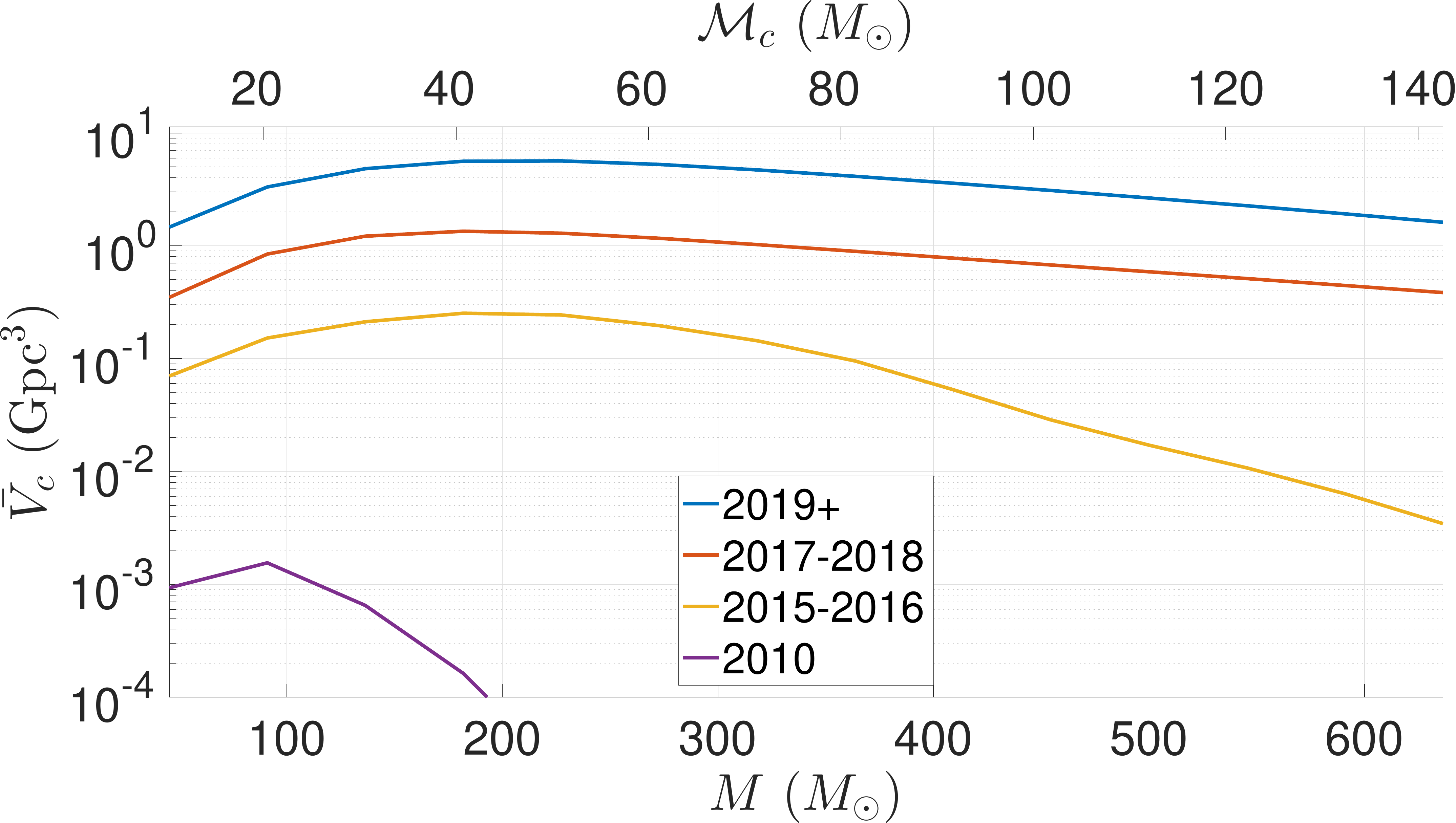}
\label{fig:sensitivity}
 \caption{Top: Horizon distance (left axis) and horizon redshift (right axis) as a function of IMBH mass for IMBH--BH coalescences with non-spinning components with a 10:1 mass ratio, for different detector sensitivities (see text). 
Bottom: Detection-weighted sensitive comoving volume, \autoref{DWSCV}; when multiplied by a constant merger rate per unit comoving volume per unit source time, this yields a detection rate.}
\end{figure*}

In \autoref{fig:sensitivity} we show the sensitivity of a network of ground-based detectors to GWs from an IMBH--BH coalescence with non-spinning components with a mass ratio of 10:1, as a function of IMBH mass. 

The top panel of \autoref{fig:sensitivity} shows the horizon distance $d_H$, which is the luminosity distance at which GWs from a face-on overhead binary would be detected at a signal-to-noise ratio of 8 by a single detector with the sensitivity of Advanced LIGO; the corresponding horizon redshift $z_H$ is shown on the right vertical axis. 
This signal-to-noise ratio is used as an approximation for sensitivity by the full network \citep{ratesdoc,GW150914:rates}; the actual sensitivity depends on the network configuration, data quality, and signal duration. 
We use the noise power spectral density (PSD) of H1 (the LIGO detector in Hanford, WA) during the S6 science run \citep[curve labeled `2010',][]{PSD:S6}, the measured noise PSD of H1 during the 2015 observing run O1 \citep[`2015--2016',][]{PSD:O1}, low-end predictions for Advanced LIGO noise PSD for the later stages of detector commissioning \cite[`2017--2018', O3 configuration of][]{scenarios}, and for design sensitivity runs in the zero detuning, high laser power configuration \citep[`2019+',][]{PSD:AL}.
We use circular effective one-body waveforms calibrated to numerical relativity for signal-to-noise-ratio calculations \citep{Taracchini:2013}.

The bottom panel of \autoref{fig:sensitivity} shows the surveyed detection-weighted comoving volume $\overline{V}_c$
\be
\label{DWSCV}
\overline{V}_c = \int_0^\infty \frac{dV_c}{dz} f_d(z) \frac{1}{1+z} dz\, ,
\ee
where $\frac{dV_c}{dz}$ is computed using the Planck \citep{Planck:2015} cosmology, $f_d(z)$ is the probability that a binary with the given source-frame masses at redshift $z$ is louder than the signal-to-noise ratio threshold of 8 (integrated over isotropically distributed sky locations and orientations), and the last factor corrects for the difference in source and observer clocks.
With this definition, $\mathcal{R} \overline{V}_c T$ yields the expected number of detections during an observing run with (at least double-coincident) duration $T$ assuming a constant merger rate $\mathcal{R}$ per unit comoving volume per unit source time.

Intermediate mass-ratio coalescences can be observed to a horizon distance of $\sim 1$ Gpc during the O1 science run, and a horizon redshift $z\sim 0.6$ at full design sensitivity. 
\autoref{fig:sensitivity} assumes a 10:1 mass ratio. 
It can be roughly rescaled to other mass ratios by noting that, for a fixed IMBH mass, the signal-to-noise ratio at a given distance, and hence the horizon distance, will scale as $\sqrt{m/M}$ when the signal is inspiral-dominated, and as $m/M$ when the signal is ringdown-dominated. 
The transition between the two regimes occurs at $M+m \sim 200\ M_\odot$ at Advanced LIGO design sensitivity \citep[see Fig.~9 of][which also discusses inference on the parameters of coalescences of intermediate mass-ratio binaries]{Haster:2015}.
As a comparison, if the BHs in GW150914 (both BHs in the initial binary and the merger product) had been the lower mass member of a 10:1 mass ratio IMBH--BH binary, all signals would be ringdown-dominated \citep{GW150914:2016}.
The same is valid for the marginally astrophysically significant event LVT151012 \citep{GW150914:rates} where only the lower mass BH from the initial binary would produce an inspiral-dominated signal.

The IMBH--BH coalescence rate is highly uncertain. 
Our simulations suggest that around one IMBH--BH merger per ten globular clusters is probable in the first $\sim$ hundred million years of the lifetime of a cluster hosting a suitable IMBH. 
The merger product may well be ejected from the cluster by the recoil kick from asymmetric GW emission in the last few pre-merger orbits. 
Assuming the IMBH is not spinning, the kick velocity for a 10:1 mass ratio coalescence is $\simeq 60 \textrm{km s}^{-1}$ \citep{Gonzalez:2007}, comparable to the typical $\sim 50 \textrm{km s}^{-1}$ escape velocity from a globular cluster. 
If so, at most $\sim$ one IMBH--BH merger would happen per cluster before the IMBH is ejected. 

Even if the merger product is retained, there is a trivial upper limit on the number of mergers per cluster in the Advanced LIGO sensitive frequency band. 
By the time the IMBH grows beyond several hundred solar masses (see \autoref{fig:sensitivity}), the sensitivity drops significantly; hence, only $\sim 30$ IMBH--BH per cluster are observable. 

Therefore, the IMBH--BH coalescence rate per suitable cluster may vary between 1 and 30 mergers over the cluster's $\sim 10$ Gyr lifetime, or $0.1$--$3$ mergers Gyr$^{-1}$. 
The space density of globular clusters is $\sim 3$ Mpc$^{-3}$ \citep{PZwart:2000}. 
Following \citet{Mandel:2008}, we will parametrize the fraction of suitable globular clusters (those with the right IMBH mass and central density) by $f$. 
Then the total merger rate is in the range $R \in [0.03 (f/0.1), (f/0.1)]$ Gpc$^{-3}$ yr$^{-1}$.
Multiplying this by the surveyed detection-weighted comoving volume, we may expect $\sim$ 0.1---5 $\times (f/0.1)$ detections per year at full sensitivity.
In the near term, 0.01---1 $\times (f/0.1)$ detections may be possible during the upcoming O2 6-month observing run, assuming a $\sim 50\%$ coincident duty cycle.\footnote{These rates should be considered in relation to the upper limits on the rate of binary IMBH coalescences from LIGO--Virgo observations, which are $\geq \mathcal{O}(10^3)$ higher depending on the IMBH masses considered \citep{LVC_IMBHB:2014}.}

This calculation may well be optimistic. 
It assumed a constant merger rate over the cluster lifetime; however, most local globular clusters are old \citep[e.g.,][]{Kruijssen:2012}, and mergers are more likely early in the cluster lifetime. 
Therefore, most mergers may happen at high redshift, where they are unlikely to be detectable. Finally, the fiducial choice $f=0.1$ is fairly arbitrary; much lower values, including $f=0$, are possible. 
On the other hand, if IMBHs are spinning, prograde inspirals could yield higher signal-to-noise ratios, and the detection volume and rate would increase even after averaging over isotropic inspiral orbits \citep{Mandel:2007}.

We have also looked for evidence of possible tidal disruptions of stars by the IMBH \citep[see][for a recent analysis]{Macleod:2016}. 
Our problem set up is not ideal for this investigation, since we have only two stellar types in addition to the IMBH: one solar mass ``stars'' and ten solar mass ``black holes''. 
None of the one-solar-mass stars in our simulation approach the IMBH within the tidal disruption radius. 
However, if we consider the black holes as proxies for evolving stars, we find that a few would approach within the IMBH tidal disruption radius while in the giant phase of their evolution. 
Given the gradual hardening of the innermost binary, it is likely that the nominal tidal disruption radius would be reached through stellar evolution rather than a dynamically-driven encounter. 
Hence, rather than transient tidal disruptions, we may expect to see Roche lobe overflows, perhaps leading to ultraluminous X-ray binaries such as ESO 243-49 HLX-1 \citep{Farrell:2009,Davis:2011,Godet:2014}.
In the case of stellar companions additional sources of apsidal precession (for example, precession due to tidal and rotational bulges) may arise and quench LK oscillations similarly to SP as discussed above. 

We have also observed a number of ejections of black holes and stars by the binary containing the IMBH.  Rapid stellar ejections, particularly at high velocities, could potentially serve as probes for the presence of an IMBH in a cluster.

The capability of our simulation tracking both the complex dynamical features and the merger stands in contrast with previous work such as \citet{Leigh:2014}, who did not include any pN effects and were thus unable to observe the quenching of LK oscillations due to SP. 
This could lead to the production of IMBH--BH binaries with artificially long lifetimes as they were unable to merge by the emission of GWs. 
This issue was partly addressed by \citet{Macleod:2016}, who included GW mergers following \citet{Peters:1964}; however, the interplay of LK effects with SP (and tides in the case of stellar companions) would likely affect their merger rate estimates.
Reaching high eccentricities through three-body interactions allows for efficient GW emission at a larger  binary semi-major axes as compared to a circular system \citep{Peters:1964,Sesana:2015}.  SP-induced freezing of the IMBH--BH orbit at high eccentricities will generally increase the probability that the binary will be highly eccentric after the next three-body interaction, thus facilitating mergers.

Finally,  the $N$-body simulations in \citet{Leigh:2014,Macleod:2016} and in this work did not include any population of primordial binaries \citep[but see][]{Heggie:2006,Trenti:2007a,Trenti:2007b}, and while binaries which does not include the IMBH can form dynamically, these would not be specifically tracked. 
The presence of primordial binary BHs in combination with an IMBH has been shown to affect the retention of BHs in the cluster as well as the evolution of the fraction of BHs in binaries \citep{Pfahl:2005,Trenti:2007a, Leigh:2014}, and would thus require a more careful treatment in future studies.

\section{Conclusion}
\label{sec:conclude}

We have, in a simulation, observed a merger of a $100 M_\odot$ IMBH and a $10 M_\odot$ BH within a globular cluster as part of the first simulation campaign accounting for post-Newtonian dynamics in the region around the IMBH. 
This has provided insight into the competitive interplay between pN effects and LK eccentricity oscillations in hierarchical systems as a mechanism for producing and hardening an IMBH--BH binary.
We have observed suppression of LK oscillations caused by the pN Schwarzschild precession of the IMBH--BH binary giving clear evidence for the necessity of including pN dynamics in future simulations of globular clusters to fully capture all relevant dynamical effects leading to the formation, evolution and merger of an IMBH--BH binary.
This is especially relevant toward the end of our simulation, where fast relativistic precession of the IMBH--BH binary freezes its orbit at high eccentricities.

Future extensions to this work will include a larger spectrum of masses (both for the IMBH and the surrounding cluster particles), longer simulation times (requiring further optimization of the code) and additional physical effects (e.g., stellar evolution, a population of primordial binaries and external tidal fields).
We have also commented on the detectability of gravitational waves emitted from an IMBH--BH merger by both space-based and ground-based observatories, with a possible detection within the next decade.

\section*{Acknowledgments}

We would like to thank Jonathan Gair, James Guillochon and Alberto Sesana for useful discussions and suggestions.
CJH acknowledges support from CIERA though a visiting pre-doctoral fellowship and a travel grant from the RAS.
FA was supported by a CIERA postdoctoral fellowship and from a NASA Fermi Grant NNX15AU69G. 
We acknowledge the use of computing resources at CIERA funded by NSF PHY-1126812.

\bibliography{IMBH_GC}

\end{document}